\documentclass[12pt,a4paper]{article}
\usepackage[all]{xy}
\usepackage{ amsthm,amsopn, graphicx, jheppub}

\newcommand{\vev}[1]{\langle #1\rangle}
\newcommand{\ket}[1]{\lvert #1\rangle}

\def\ket#1{\langle #1 \rangle}

\newcommand{\nn}{\nonumber}

\def\beq{\begin{equation}}
\def\eeq{\end{equation}}
\def\beqa{\begin{eqnarray}}
\def\eeqa{\end{eqnarray}}

\newcommand{\PP}{\mathbb{P}}
\newcommand{\CC}{\mathbb{C}}
\newcommand{\XX}{$\mathcal{X}$}
\newcommand{\A}{$\mathcal{A}$}

\DeclareMathOperator{\Conf}{Conf}
\DeclareMathOperator{\Gr}{Gr}

\def\matt[#1,#2,#3,#4]{\left(%
\begin{array}{cc} #1 & #2 \\ #3 & #4 \end{array} \right)}

\def\v2#1{\vv2[#1]}
\def\vv2[#1,#2]{\left(%
{#1 \atop #2}\right)}

\def\nn{\nonumber}

\preprint{IPhT}
\keywords{QCD, gauge symmetry, scattering amplitudes}
\title{Cluster algebras in scattering amplitudes with special 2D kinematics}
\author{Marcus A. C. Torres}
\affiliation{Institut de Physique Th\'eorique, CEA-Saclay, F-91191 Gif-sur-Yvette
cedex, France\\
{\rm email: marcus-andre.de-carvalho-torres@cea.fr}}

\abstract{ We study  the cluster algebra of the kinematic configuration space $\Conf_n(\PP^3)$ of a $n$-particle scattering amplitude restricted to the special 2D kinematics. We found that the $n$-points two-loop MHV remainder function in special 2D kinematics depend on a selection of \XX-coordinates that are part of a special structure of the cluster algebra related to snake triangulations of polygons. This structure forms a necklace of hypercubes beads in the corresponding Stasheff polytope. Furthermore at $n = 12$, the cluster algebra and the selection of \XX-coordinates in special 2D kinematics replicates  the cluster algebra and the selection of \XX-coordinates of  $n=6$ two-loop MHV amplitude in 4D kinematics.}

\begin{document}
\maketitle


\section{Introduction}

  $\mathcal{N}=4$ supersymmetric Yang-Mills in its planar limit is the present arena for developing new methods of calculating scattering amplitudes of perturbative QCD.  Feynman diagrams revealed itself to be cumbersome and incapable of unveiling the beautiful symmetries and structures of SYM. 

Among these symmetries, it was realized \cite{Drummond:2006rz} that $\mathcal{N}=4$ SYM at tree level is superconformal in Minkowski space and in the dual space and at any loop order one can calculate amplitudes by  calculating Wilson loops in the dual space, which led to several results\cite{am,dks,bht,dhks4,Drummond:2007au,seven,dhks6,Anastasiou:2009kn}.

One of the present challenges of SYM is that amplitudes at  L-loops involve complicated polylogarithm functions of transcendentally 2L.  Such functions have several relations among them and an amplitude can be written in different forms. Also there are numerous choices of conformal cross ratios as arguments.
  
In \cite{Goncharov:2010jf,Golden:2013xva}, the authors Golden, Goncharov, Spradlin, Vergu and Volovich (GGSVV) showed that a judicious choice of kinematic variables was one of the main ingredients in a large simplification of the previously calculated two-loops six particle MHV remainder function $\mathcal{R}_n^{(2)}$  \cite{DelDuca:2009au,DelDuca:2010zg,Zhang:2010tr} and that this choice is related to the cluster structure that is intrinsic to the kinematic configuration space $\Conf_n(\PP^3)$ of external particles.  Its cluster structure selects the appropriate cross ratios (directly related to $\mathcal{X}$-coordinates in the cluster algebras) to be used in the motivic amplitudes \cite{Golden:2013xva} and intriguingly some cluster algebras define arguments so suitably to some polylogarithmic functional equation as the famous Abel's pentagon dilogarithm identity and a recently found trilogarithm relation \cite{Golden:2013xva}, showing that the use of cluster coordinates as arguments in remainder functions may be the appropriate way to simplify the long logarithm expressions. Fortunately, this cluster structure that belongs to the kinematic configuration space is shown  \cite{Golden:2013xva} to be preserved at two-loop order where weight four polylogarithmic identities are less known. Other interesting properties of the cluster structure of the configuration space, such as positivity and the fact that the logarithms of their $\mathcal{X}$-coordinates are canonical Darboux coordinates of a Poisson space \cite{clps}, led the authors of \cite{Golden:2013xva} to propose that the variables which should appear in the motivic MHV amplitudes in SYM are  cluster  $\mathcal{X}$-coordinates of the cluster algebra of its configuration space. 

The use of the cluster structure of  $\Conf_n(\PP^3)$ worked efficiently for $n=6, 7$, but for $n$ greater than 7, the cluster algebra is of infinite type and we cannot count all of its \A-coordinates, \XX-coordinates and clusters. In order to overcome such problem, here we focus on the study of  finite cluster sub-algebras for $n\ge 8$ by reducing the configuration space to a certain kinematic limit.

In special 2D kinematics, the cluster sub-algebras are always finite and in this case the algebra can be related to a Lie Algebra with simply laced Dynkin diagram \cite{Fomin}. Also the two-loop MHV remainder function in special 2D kinematics  has been fully calculated at $n$ larger than 7 \cite{Heslop:2010kq} and we are able to check the existence of the cluster structure in amplitudes in this kinematic domain.

The program of research initiated by \cite{Golden:2013xva} is a very instigating one and many questions remain to be answered. For example, at $n=$ 6 and at $n=$  7 only $3/5$ of all cluster \XX-coordinates of their respective cluster algebra show up in the two-loop MHV motivic amplitude. Another question is that present studies of cluster algebras are done concerning results on two-loop MHV  amplitudes only \cite{Golden:2013lha}, but  possibly some of the cluster structure of the configuration space is preserved at higher loops and non MHV amplitudes.

 In this regard, a small drawback exist in special 2D kinematics concerning its cluster structure. At $n=8$,  ${\rm N}^2{\rm MHV}$ one-loop, NMHV two-loop and MHV three-loop  amplitudes (remainder functions) were calculated in \cite{Caron-Huot:2013vda} and these results were found to have its symbols, polylogarithm arguments and poles with mixing \XX-coordinates v and w of its $A_1\times A_1$ cluster algebra. The mixing of \XX-coordinates occur in the form of $v-w$ and $1-vw$.  This structure is believed to exist in 2D special kinematics for $ k+l\ge 3$ ${\rm N}^k{\rm MHV}$ l-loop amplitudes. Nevertheless, the cluster structure continue to have a role on the selection of cross ratios ($v$ and $w$). The different structure on the arrangement of \XX-coordinates in arguments of polylogarithm expressions in special 2D kinematics compared with two-loop MHV amplitudes in 4D kinematics may be connected to the fact that the positivity condition \footnote{Observed by Song He in private communication.} of cluster \A-coordinates  in special 2D kinematics and 4D kinematics are not the same while the amplitude in special 2D kinematics is a reduction from the amplitude in 4D kinematics.  

 In addressing the study in a smaller configuration space we were also able to recognize that not all \XX-coordinates appear in the remainder function and we were able to identify their main characteristics within the cluster structure, through the use of associated polygon triangulations and Stasheff polytopes. We were able to find in special 2D kinematics at $n=12$, a double copy of the same $A_3$ cluster algebra that appears in the configuration space of $n=6$ in 4D kinematics and equally the same 9 out of 15 (for each copy) of the \XX-coordinates appear in the remainder function. These  9 \XX-coordinates are sorted as \XX-coordinates of the 6 snake clusters of $A_3$ cluster algebra. We call snake clusters, the ones whose associated polygon triangulation has their diagonals associated to negative simple roots \cite{FominZ}. Such triangulations have a zig-zag or snake appearance that name them.   We notice that snake clusters are part of a structure in their corresponding Stasheff polytope. We call it necklace of hypercube beads, or hypercube necklace. In this structure (necklace) in the Stasheff polytope, the nearest snake cluster vertices are opposite vertices of a hypercube that connects them. All quadrilateral faces ($A_1\times A_1$ cluster subalgebras) in these hypercubes have its two \XX-coordinates belonging to snake clusters. These snake clusters are the two which correspond to the vertices that connect the hypercube to the necklace in the Stasheff polytope.
 
 In \cite{Golden:2013xva}, the authors related  obstruction terms in the amplitude expression to quadrilateral faces in the Stasheff polytope. Such terms are obstructions to write the two-loop MHV motivic amplitude in terms of classical 4-logarithms. In special 2D kinematics, such motivic amplitudes are trivial since the coproduct of product of logarithms are trivial. Consequently, obstructions cannot be studied here but the cluster algebra that we study here is present in 4D kinematics where quadrilateral faces with selected snake cluster \XX-coordinates may play such role.
 
  Our hope is that a similar study of \XX-coordinates may exist for type  $E_6$ cluster algebras or its cluster sub-algebras, which may help justify in $n= 7$ \cite{Golden:2013xva} why only $3/5$ of all \XX-coordinates of $E_6$ cluster algebra are used in the amplitude and the use by the obstruction terms of only 42 quadrilateral faces out of 1785 quadrilateral faces existent in $E_6$ Stasheff polytope.  

This present work is divided as follow. In section \ref{2Dkinematics}, we present special 2D kinematics \cite{Heslop:2010kq,Goddard:2012cx} and cross ratios. In section \ref{cluster}, we review elements of cluster algebra. In section \ref{confspace}, we reduce the configuration space to special 2D kinematics and find its cluster algebra, its cluster coordinates and mutation relations. In section \ref{results}, we study the cases for n = 8, 10 and 12. The cases n = 8 and 10  show the presence of the cluster structure of the configuration space among the arguments  (cross ratios) of $\mathcal{R}_n^{(2)}$ in special 2D kinematics while at n = 12, we notice a relevant structure related to a selection of \XX-coordinates by the amplitude which is further studied in the general case of n external particles in section \ref{general n}. We conclude in section \ref{conclusion}.  

\section{Special 2D kinematics}
\label{2Dkinematics}

 A good description of special 2D kinematics is presented in \cite{Heslop:2010kq, Goddard:2012cx} and we review a few aspects below.

Special two-dimensional kinematics is the  condition when all external particles have their momenta lying in the same $1+1$ dimensions while internal particles propagate in $3+1$ dimensions. The external momenta $p_i$ define a polygon in the dual space $x^{\mu}$ (up to global translation), 
\beq
p_i^{\mu}=x_i^{\mu}-x_{i+1}^{\mu}
\eeq
 
The edges of the polygon, $p_i$, zig-zag inverting  spacial direction, i. e., switching between light cone directions $x^+$ and $x^-$ \cite{Alday:2009yn} and this condition implies that there must be an even number of external particles.

As a result, the external momenta of particles in light cone $(+,-)$ directions are:
\begin{align}\label{eq:6}
  p_{i}=\left\{
  \begin{array}{ll}
( p_{i}^+,0)\ , \qquad &i\ \rm{even}\\
       ( 0,p_{i}^-)\ , \qquad &i\ \rm{odd}
     \end{array}\right.
\end{align}
In the language of momentum twistors, it translates into:
\begin{align}
\label{2dtwist}
  Z_{i}=\left\{
  \begin{array}{ll}
(Z_i^{1},0,Z_i^{3},0) &\qquad i\ \rm{even}\\
(0,Z_i^{2},0,Z_i^{4}) &\qquad i\ \rm{odd}\ ,
\end{array}
\right.
\end{align}
 reducing $SL(4) \to SL(2)_+ \times SL(2)_-$ in 2d.
The $SL(4)$ invariant product of four momentum twistors 
\begin{align}\label{eq:3}
  \vev{ijkl}=\epsilon_{abcd} Z_i^a Z_j^b Z_k^c Z_l^d\ ,
\end{align}
becomes zero, unless there are two odd and two even indices. In this case the even and odd indices factorize into $SL(2)_{\pm}$ invariant terms. For example, 
 \begin{align}\label{eq:4}
  \vev{1234}=\vev{13}\vev{24} = (Z_1^{2}Z_3^{4}-Z_1^{4}Z_3^{2})(Z_2^{1}Z_4^{3}-Z_2^{3}Z_4^{1})
\end{align}

Writing Lorenz invariant square distances of dual coordinates in terms of momentum twistors \cite{ArkaniHamed:2010gh}:
\beq
(x_{ij})^2=(x_i-x_j)^2= \frac{\vev{ii+1jj+1}}{\vev{\lambda_i\lambda_{i+1}}\vev{\lambda_j\lambda_{j+1}}}
\eeq

and applying  it in a standard basis of cross ratios  in 4 dimensional kinematics  \cite{Heslop:2010kq}
\beq
 u_{ij}=\frac{x_{ij+1}^2 x^2_{i+1 j}}{ x_{ij}^2 x_{i+1 j+1}^2} =\left\{
  \begin{array}{ll}
1&\qquad i\ \rm{and}\ j\ \rm{with\, opposite\, parity} \\
\frac{\vev{ij+2}\vev{i+2j}}{\vev{ij}\vev{i+2j+2}} &\qquad i\ \rm{and} \ j\ \rm{with\, same\, parity}
\label{crossratios}
\end{array}
\right.
\eeq

Therefore, in 2D kinematics the cross ratios are separate in two groups: those that depend only on  momentum twistors with even indices and those that depend only on momentum twistors with odd indices.

In order to keep notation in agreement with \cite{Heslop:2010kq}, we rewrite the above 2D cross ratios as:
\begin{equation}\label{uijdef2}
  u_{ij}^{+}\, :=\,\frac{\vev{2i+1,2j+3}\vev{2i+3, 2j+1}}{\vev{2i+1,2j+1}\vev{2i+3, 2j+3}}  \ , \qquad
   u_{ij}^{-}\, :=\,\frac{\vev{2i,2j+2}\vev{2i+2, 2j}}{\vev{2i, 2j}\vev{2i+2, 2j+2}}  \ ,
\end{equation}

Applying a series of Pl\"ucker identities
\beq
\vev{ij}\vev{kl}=\vev{ik}\vev{jl}+\vev{il}\vev{kj}
\eeq
we can check the Y-system (evaluated at fixed spectrum parameter $\zeta=0$) found in \cite{Heslop:2010kq, Alday:2009dv}:
\begin{align}
\label{ueq}
  (1-u^\pm_{i\, j+1})   (1-u^\pm_{i+1\, j})& \,=\,   (1- 1/{u^\pm_{i\,
      j}})  (1-1/{u^\pm_{i+1\, j+1}})
\end{align}
which constitute in two separate set of equations keeping the sets of cross ratios ${u^+}$ and ${u^-}$ independent from each other. 

\section{Cluster algebras and Stasheff polytopes}
\label{cluster}

The subject of cluster algebras was recently presented in \cite{Golden:2013xva}, in a accessible way to physicists and complementing the standard references \cite{1057.53064,1215.16012, Fomin, FominZ}.  We  review here only basic concepts and  terms.

We are only interested in the finite type cluster algebras.  Such cluster algebras have a finite set of distinct  generators (cluster variables), that is grouped in a finite number of clusters (sets) of equal size and that relate to each other by exchange relations where one of the cluster variables is replaced by (mutates to)  another cluster variable outside the cluster. These exchange relations  can be codified within each cluster by associating them with oriented quivers. From the quiver associated to a cluster we can define \XX-coordinates related to each cluster variable in that cluster.   

A cluster may contain a subset of frozen variables (or coefficients)  that do not mutate and stays the same in all clusters. The number of  cluster variables (not frozen ones) in a cluster is the rank of the cluster algebra. We call both cluster variables and cluster coefficients as \A-coordinates.  

Quivers are built with arrows connecting vertices. Such vertices in a quiver are identified with \A-coordinates while arrows  define exchange relations among \A-coordinates and the \XX-coordinates of each vertex in the quiver.  Quivers are such that loops and two-cycles are not allowed. Loops are arrows that have same origin and target and two-cycles are a pair of arrows with opposite direction connecting the same two vertices. When a two-cycle appears after a mutation, the arrows ``cancel each other'' and disappear.

A mutation of a cluster variable in a cluster, or vertex $k$ in the corresponding quiver transforms it to a new quiver according to the following operations:
\begin{itemize}
\item for every pair of arrows $i\rightarrow k$ and $k\rightarrow j$, add a new arrow $i\rightarrow j$,
\item reverse all arrows that target k or depart from k,
\item proceed with all two-cycle cancelation.
\end{itemize}

A theorem \cite{FominZ}  classifies all finite type cluster algebras according to simply  laced Lie Algebras. It states that given a  finite type cluster algebra, their clusters have quivers  that are mutation equivalent to a Dynkin diagram of a Lie Algebra, via identification of the principal part of its quiver. The principal part of a quiver is the quiver without frozen variables and arrows to or from them. 

There can be more than 1 arrow between 2 vertices and a number can be added on top of each arrow for cases of multiple arrows. An skew symmetric adjacency matrix $(b_{ij})$ can be defined from the quiver, where
\beq
b_{ij}= \# arrows (i\rightarrow j)-\#arrows (j\rightarrow i)
\eeq

A cluster variable in vertex $k$, $a_k$, mutates to $a'_k$ according to the exchange relation:
\beq
a'_k a_k= \prod_{i| b_{ki}>0}a_i^{b_{ki}}+\prod_{i| b_{ki}<0}a_i^{-b_{ki}}
\eeq 
 
 In a quiver, for every vertex  correspondent to a cluster variable we define its \XX-coordinate $x_i$ in terms of \A-coordinates $a_j$,
\beq
x_i= \prod_{j\neq i}a_j^{b_{ij}}
\label{xdef}
\eeq

We remark that when a mutation occurs in one vertex forming a new cluster, the adjacency matrix changes accordingly and following (\ref{xdef}) the new cluster will have different \XX-coordinates. The \XX-coordinate of vertex $i$ under mutation, mutates from $x_i$ to $x_i^{-1}$. Throughout the paper, we will not count a \XX-coordinate  and its inverse as independent \XX-coordinates.

A useful construction associated to a finite type cluster algebra is a generalized associahedron. Such construction represents the cluster algebra as a polytope with clusters being represented by vertices and mutations between clusters being represented by edges connecting vertices. For a rank $r$ cluster algebra, each vertex is parametrized by the $r$ \XX-coordinates of the represented cluster and from each vertex departs $r$ edges. 

 In type A cluster algebra  the generalized associahedron is called Stasheff polytope \cite{FominZ}, which will be used from now on, since we will be always dealing with type A cluster algebras in special 2D kinematics.

Naming the rank of a Stasheff polytope as the rank of the corresponding cluster algebra, an interesting property of a Stasheff polytope  is that lower rank Stasheff polytopes corresponding to cluster subalgebras can be easily identified in the polytope. 

Rank one $A_1$  and rank two  $A_1\times A_1$ and $A_2$ cluster algebras are associated to the smallest Stasheff polytope which are dimension one edge and dimension two quadrilateral and pentagonal faces, respectively.

\section{Configuration space and its cluster structure}
\label{confspace}

 A configuration space is the space of parametrization of amplitudes. In four dimensions, the configuration space has dimension $(3n-15)$ as the space of external $n>4$ points in  $\PP^3$ (momentum twistors) modulo the action of the conformal group $PGL_4$ in $\PP^3$. We denote it by $\Conf_{n}(\PP^3)$. It can also be interpreted as the space of $4\times n$ matrices quotient by the conformal group SL(4) and with each column quotient by $\CC^*$ rescaling . The quotient of the space of $4\times n$ matrices  by the action of SL(4) and the diagonal subgroup $\CC_{diag}^*$ of $(\CC^*)^n$ (which rescales all columns by the same factor) is the Grassmannian  $\Gr(4,n)$. Therefore,

\beq
\Conf_{n}(\PP^3)\cong \Gr(4,n)/ (\CC^*)^{n-1}
\eeq
  
 In the case of special 2D kinematics, momentum twistors (\ref{2dtwist}) have one degree of freedom each and there are 6 conformal symmetries which reduces the  configuration space to $(n-6)$ dimensions. 

As reviewed above, the cross ratios  $u_{ij}^{\pm}$ depend only on either even or odd momentum twistors and they have separate equations (\ref{ueq}) which leads us to expect that the configuration space to be divided in two subspaces of $(n/2-3)$ dimensions each.

 Momentum twistors contain only 2 nonzero components in special 2D kinematics. Considering their little group $\CC^*$ each momentum twistor can be seen as a point in $\PP^1$. Two separate $(2\times n/2)$ matrices can be constructed out of $n/2$ points (even or odd indices only) in $\PP^1$. They transform with the action of the conformal groups (in special 2D kinematics) $SL(2)_+$ and $SL(2)_-$, respectively. Hence considering these two matrices modulo $SL(2)_+ \times SL(2)_-$ action and modulo a diagonal $\CC_{diag}^*$ of each $(\CC^*)^{n/2}$ , the configuration space is equivalent to two Grassmannians $\Gr(2,n/2)_+\times\Gr(2,n/2)_-$ of even and  odd momentum twistors respectively, each  modulo the action of the remaining little group $(\CC^*)^{n/2-1}$ of the momentum twistors. We write it as $\Conf_{n/2}(\PP^1)\times\Conf_{n/2}(\PP^1)$ and each copy $\Conf_{n/2}(\PP^1)$ has dimension $n/2-3$.

Our remaining task is to study the cluster algebra of $\Gr(2,n/2)$. The initial quiver \cite{1057.53064, 1215.16012} can be drawn as follow
\begin{eqnarray}
\begin{gathered}
\begin{xy} 0;<1pt,0pt>:<0pt,-1pt>::
(25,25) *+{\langle 2, n/2\rangle} ="0", 
(100,25) *+{\langle 2, n/2-1\rangle}="9",
(160,25)*+{\; . \;.\; .\; }="7",
(200,25) *+{\langle 2 4\rangle} ="1",
(250,25) *+{\framebox[5ex]{$\langle 2 3\rangle$}} ="2",
(250,75) *+{\framebox[5ex]{$\langle34\rangle$}} ="3",
(200,75) *+{\framebox[5ex]{$\langle45\rangle$}} ="4",
(160,75) *+{\; . \;.\; .\; }="8",
(100,75) *+{\framebox[13ex]{$\langle n/2\hspace{-1mm}-\hspace{-1mm}1, n/2\rangle$}}="10",
(25,75) *+{\framebox[9ex]{$\langle 1, n/2\rangle$}} ="5",
(-5,-5) *+{\framebox[5ex]{$\langle 12\rangle$}} ="6",
(160,75) *+{},
"0", {\ar"9"},
"9",{\ar"7"},
"9",{\ar"10"},
"7", {\ar"1"},
"8", {\ar"9"},
"10", {\ar"0"},
"0", {\ar"5"},
"6", {\ar"0"},
"1", {\ar"2"},
"3", {\ar"1"},
"1", {\ar"4"},
\end{xy}
\end{gathered}
\label{initial cluster}
\end{eqnarray}

 Here the numbers {i} in $\langle i...\rangle$ are column positions in a general $\Gr(2,n/2)$ and they correspond to momentum twistor position-index 2i in $\Gr(2,n/2)_+$ and position-index 2i-1 in $\Gr(2,n/2)_+$ case. For the rest of this section we will work on a general $\Gr(2,n/2)$, except when mentioned otherwise.
 
This quiver has the form of the $A_{n/2-3}$ Dynkin diagram which classifies it as a finite cluster algebra of type $A_{n/2-3}$ \cite{FominZ} . The rank  of the Lie algebra corresponds to the number of cluster variables in a cluster, and is equal to the dimension of  $\Conf_{n/2}(\PP^1)$. The positive roots of the Lie algebra is in bijective correspondence with the non-initial cluster variables. In  $A_m$, there are $m(m+1)/2$ positive roots.  Another important fact from $A_m$ cluster algebras is that the number of clusters is given by the number of triangulations of a polygon with $(m+3)$ sides and it is the $(m+1)$th Catalan number $C_{m+1}$.

In summary, this is what we can say about a $A_{n/2-3}$ cluster algebra:
\begin{itemize}

\item 
 Its clusters (or quivers) contain $n/2-3$ cluster variables each. 
 \item The total number of cluster variables is  equal the rank $(n/2-3)$ plus the number of positive roots of $A_{n/2-3}$ root system:
 \beq
 \frac{(n/2-3)n}{4}
\label{Aalgebra} 
 \eeq
\item The number of clusters is 
$\quad C_{n/2-2}=  \frac{(n-4)!}{(n/2-1)!(n/2-2)!}$  
\end{itemize} 

The $\mathcal{X}$-coordinates of the initial quiver can be easily worked out from (\ref{initial cluster}). For $4\leq j < n/2$, the $\mathcal{X}$-coordinates corresponding to $\mathcal{A}$-coordinates $\langle 2,j\rangle$ is:
\beq
\frac{\langle 2,j-1\rangle\langle j,j+1\rangle}{\langle j-1,j\rangle\langle 2,j+1\rangle}\quad {\rm or\; (for\; j=n/2)} \quad \frac{\langle 1,n/2\rangle\langle 2,n/2-1\rangle}{\langle 12\rangle\langle n/2-1,n/2\rangle} 
\label{initialcoords}
\eeq

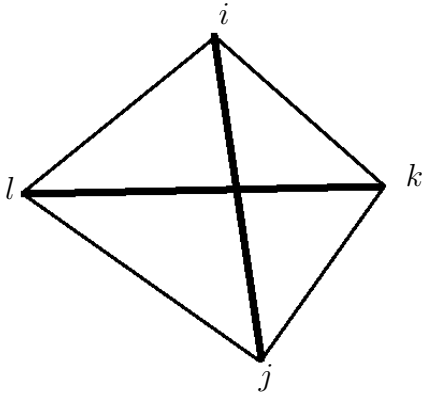
\begin{figure}
\begin{picture}(150,150)(-80,0)
\put(0,0){\includegraphics[scale=0.4]{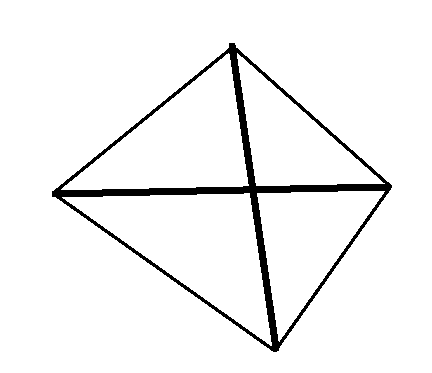}}
\put(95,136){$i$}
\put(165,75){$k$}
 \put(110,0){$j$}
 \put(15,70){$l$}
\end{picture}
\caption[diagonals]{ Diagonals $\overline {ij}$ and $\overline {kl}$ define two possible triangulations for the quadrilateral ikjl.}
\label{triangulation}
\end{figure}

We identify all cluster coordinates from a $A_{n/2-3}$ cluster algebra, via a geometric interpretation of the cluster algebra as triangulations of a polygon \cite{FominZ, 1215.16012} in the following way:
\begin{itemize}
\item a cluster is associated to a triangulation of a $n/2$-gon, such that no diagonal cross one another.
\item \A-coordinates $\vev{ij}$ (with $i < j$) of a cluster correspond to edges linking vertices i and j of the polygon. They are frozen variables when they correspond to side edges ($j=i+1$) and cluster variables when correspond to diagonals ($j>i+1$) of the polygon triangulation.
\item  a mutation is associate to changing one diagonal to another diagonal such that both are diagonals of the same quadrilateral (figure \ref{triangulation}).
\end{itemize}

We can also identify the \XX-coordinates of a given cluster by the associated polygon triangulation.
For $i<k<j<l$, the diagonals of figure (\ref{triangulation})  represent two \A-coordinates $\vev{ij}$ and $\vev{kl}$, one is a  mutation from another. Their \XX-coordinates are given by the sides of the quadrilateral ikjl:
\beq
x_{ij}=x_{kl}^{-1}=\frac{\vev{ik}\vev{jl}}{\vev{kj}\vev{il}}
\label{ijcondition}
\eeq

\begin{figure}
   \begin{center}
  \includegraphics[scale=0.35]{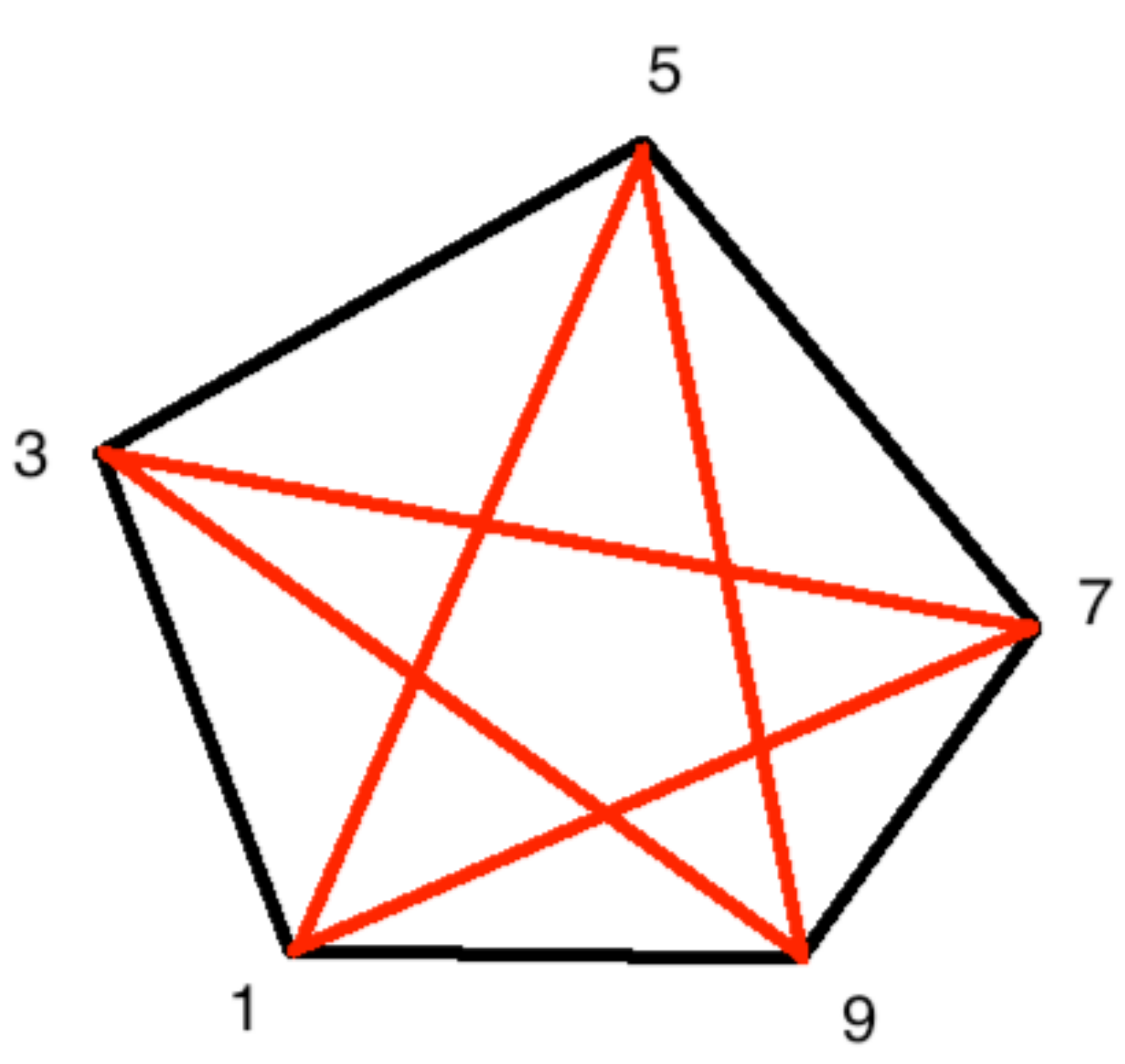}
  \hspace{1cm}
  \includegraphics[scale=0.35]{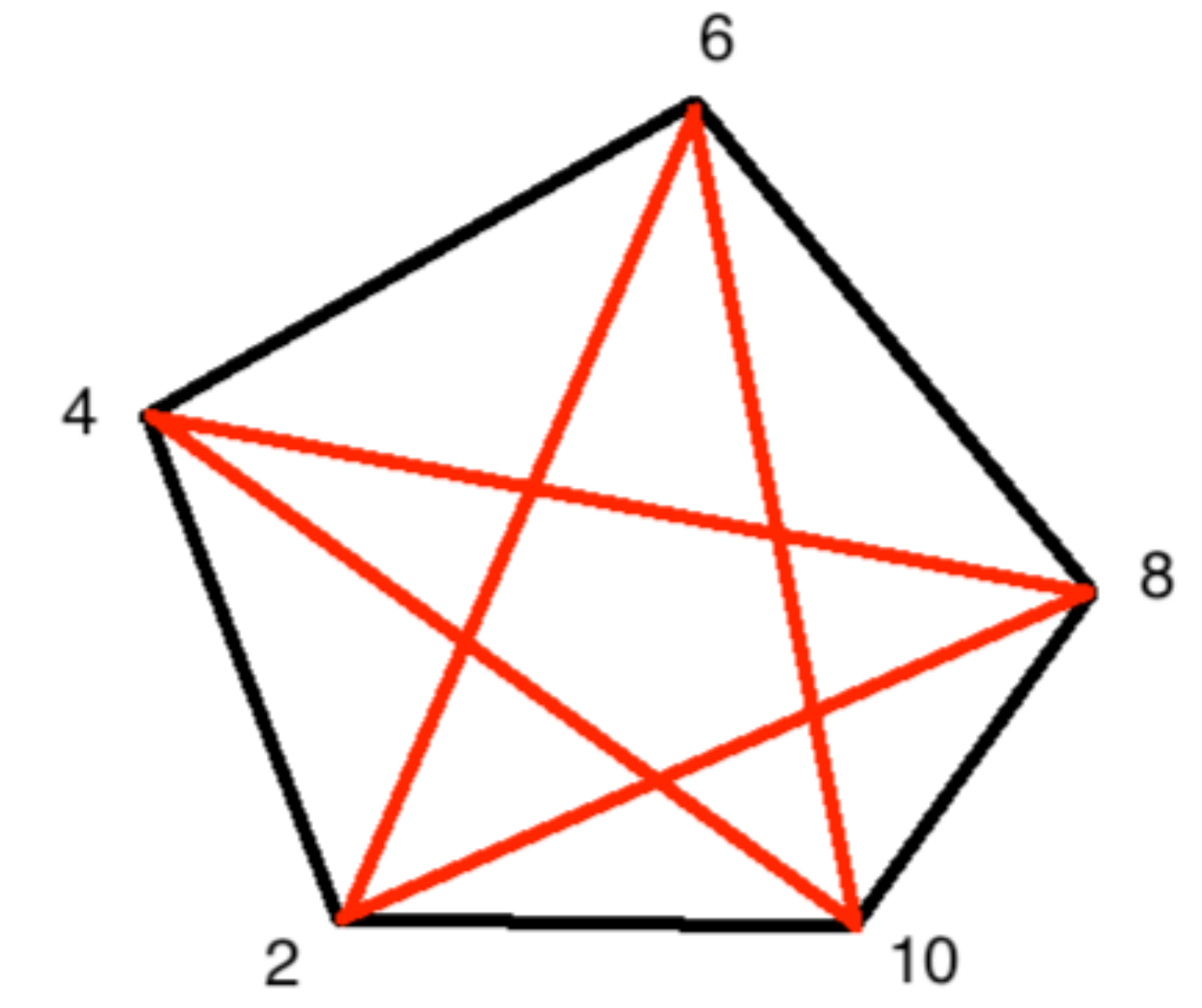}
  \end{center}
  \caption[Pentagram]{Pentagons corresponding to a $A_2\times A_2$ cluster algebra for $n=10$ external particles in 2D kinematics. Their ten diagonals correspond to all cluster variables.}
  \label{A2A2}
\end{figure}

   In order to  $\mathcal{A}$-coordinates $\vev{ij}$ of a $\Gr(2,n/2)_{\pm}$ cluster sub-algebra  to be associated to diagonals (and not sides)  in a $n/2$-gon    whose vertices are all even or all odd (figure \ref{A2A2}), it is required that for $i<j$, 
   \beq
   4\leq j-i \leq n-4
   \label{conditions}
   \eeq   
   
\section{Results for $n = 8, 10$ and $12$}
\label{results}

For such cases we are able to directly compare cluster coordinates with the cross ratios found in the two-loop MHV remainder functions expressed in \cite{Heslop:2010kq}. These cases show how cross ratios fit as \XX-coordinates. In the third case ($n=12$) we observe a special selection of \XX-coordinates which we will explore more and generalize in the next section. 

\subsection{$n=8$}

 At $n=8$ the $A_1\times A_1$ cluster algebra is of rank 2 and there are  2  clusters for each $A_1$ cluster sub-algebra making a total of 4 cluster \A-coordinates.  Equation (\ref{initialcoords}) determines  two initial \XX-coordinates:
  \beq
x^+= \frac{\langle 1,7\rangle\langle 3,5\rangle}{\langle 1,3\rangle\langle 5,7\rangle}\quad {\rm and} \quad x^-=\frac{\langle 2,8\rangle\langle 4,6\rangle}{\langle 2,4\rangle\langle 6,8\rangle} 
\eeq

Using the nomenclature in (\ref{crossratios}) and (\ref{uijdef2}), we find that:

\begin{eqnarray}
\label{eightcrosses}
  \quad
 u_{15}=u^+_{24} = \frac{x^{+}}{ 1 + x^{+}}= \frac{1}{ 1 +\frac{ 1}{ x^{+}}} &\ ,& \qquad u_{26} =u^-_{13} = \frac {x^{-}}{ 1 + x^{-}}= \frac{1}{1 +\frac{ 1}{ x^{-}}}
\ ,
\\ 
 u_{37}=u^+_{13} = \frac{1}{ 1 + x^{+}}&\ ,& \qquad u_{48}=u^-_{24}  = \frac {1}{ 1 + x^{-}}\ ,
\\ \nonumber
\end{eqnarray}

 From above, $x^+$ and $x^-$ are the well known cross ratios $\chi^+$ and $\chi^-$ \cite{Alday:2009yn, Brandhuber:2009da, Heslop:2010kq}. The remaining cluster $\mathcal{X}$-coordinates are their respective cluster mutation  $(x^+)^{-1}$ and $(x^-)^{-1}$. The standard cross ratios $u_{ij}$ are  simple transformation of the cluster  $\mathcal{X}$-coordinates, as remarked in  \cite{Golden:2013xva, Golden:2013lha}:
\beq
u_i= \frac{1}{1+v_i}
\label{simplefunction}
\eeq 

where  $u_i$ is a cross ratio and $v_i$ is a $\mathcal{X}$-coordinate. Such transformation preserves the intrinsic positivity of $\mathcal{X}$-coordinates, essential to keep cross ratios in good kinematic domain.  See \cite{Golden:2013xva} for discussion on good kinematic domain.

\subsection{$n=10$}

The corresponding $A_2\times A_2$ cluster algebra is of rank 4 and there are 10 cluster variables and 10 \XX-coordinates. Equation (\ref{initialcoords}) determines the four initials \XX-coordinates:
\beq
\left\{\frac{\vev{3,7}\vev{1,9}}{\vev{1,3}\vev{7,9}},\frac{\vev{3,5}\vev{7,9}}{\vev{3,9}\vev{5,7}}\right\}\quad {\rm and}\quad \left\{\frac{\vev{4,8}\vev{2,10}}{\vev{2,4}\vev{8,10}},\frac{\vev{4,6}\vev{8,10}}{\vev{4,10}\vev{6,8}}\right\}
\eeq

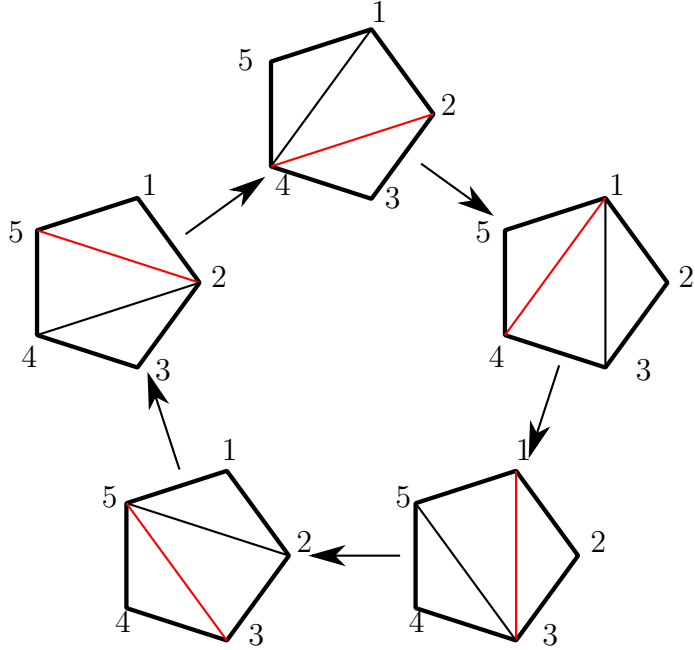
\begin{figure}
\centering
\begin{picture}(200,200)
\put(0,0){\includegraphics{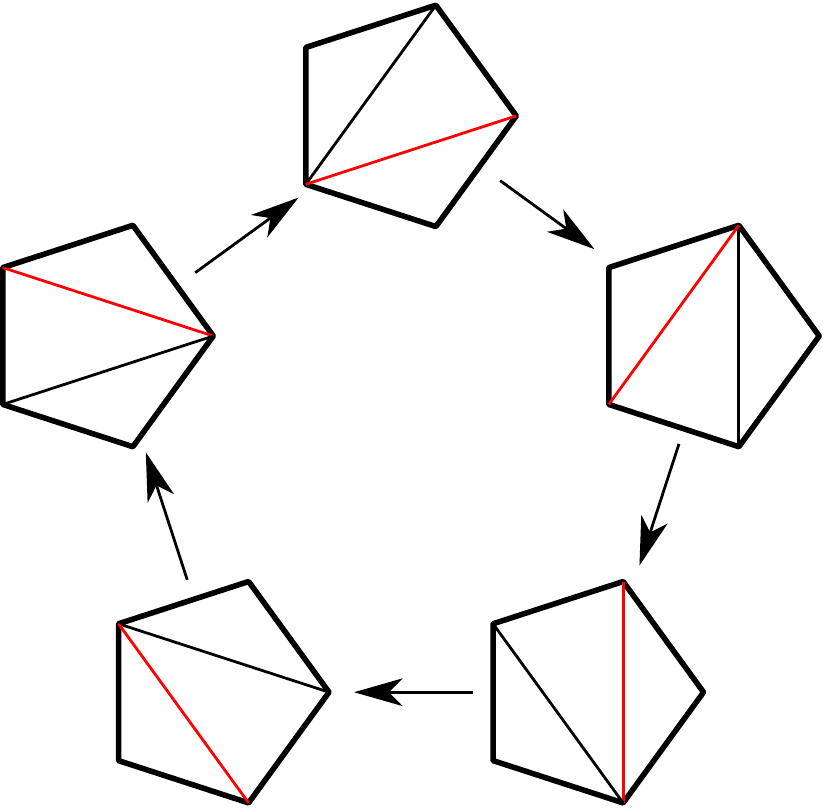}}
\put(40,170){$1$}
\put(66,135){$2$}
 \put(45,100){$3$}
 \put(-5,105){$4$}
 \put(-10,152){$5$}
 \put(126,235){$1$}
\put(152,200){$2$}
 \put(131,165){$3$}
 \put(90,170){$4$}
 \put(76,217){$5$}
\put(215,170){$1$}
\put(241,135){$2$}
 \put(225,100){$3$}
 \put(170,105){$4$}
 \put(165,152){$5$}
\put(70,70){$1$}
\put(98,35){$2$}
 \put(80,0){$3$}
 \put(30,5){$4$}
 \put(25,52){$5$}
 \put(180,70){$1$}
\put(208,35){$2$}
 \put(190,0){$3$}
 \put(140,5){$4$}
 \put(135,52){$5$}
\end{picture}
\caption[Pentagon]{Transition between pentagon triangulations representing transition between different clusters via mutation of one of its coordinates (red diagonal).}
\label{5transition}
\end{figure}

The remaining \XX-coordinates can be retrieved by  following  transitions of pentagon triangulations (figure \ref{5transition})  which represent cluster transitions by mutating one coordinate at a time:

\begin{eqnarray}
\left\{\frac{\vev{1,3}\vev{7,9}}{\vev{3,7}\vev{1,9}},\frac{\vev{1,7}\vev{3,5}}{\vev{1,3}\vev{5,7}}\right\}\quad {\rm ,}\quad \left\{\frac{\vev{2,4}\vev{8,10}}{\vev{4,8}\vev{2,10}},\frac{\vev{2,8}\vev{4,6}}{\vev{2,4}\vev{6,8}}\right\}\\ \nn
\left\{\frac{\vev{1,5}\vev{7,9}}{\vev{1,9}\vev{5,7}},\frac{\vev{1,3}\vev{5,7}}{\vev{1,7}\vev{3,5}}\right\}\quad {\rm ,}\quad \left\{\frac{\vev{2,6}\vev{8,10}}{\vev{2,10}\vev{6,8}},\frac{\vev{2,4}\vev{6,8}}{\vev{2,8}\vev{4,6}}\right\}\\ \nn
\left\{\frac{\vev{1,9}\vev{5,7}}{\vev{1,5}\vev{7,9}},\frac{\vev{1,3}\vev{5,9}}{\vev{1,9}\vev{3,5}}\right\}\quad {\rm ,}\quad \left\{\frac{\vev{2,10}\vev{6,8}}{\vev{2,6}\vev{8,10}},\frac{\vev{2,4}\vev{6,10}}{\vev{2,10}\vev{4,6}}\right\}\\ \nn
\left\{\frac{\vev{5,7}\vev{3,9}}{\vev{3,5}\vev{7,9}},\frac{\vev{1,9}\vev{3,5}}{\vev{1,3}\vev{5,9}}\right\}\quad {\rm ,}\quad \left\{\frac{\vev{6,8}\vev{4,10}}{\vev{4,6}\vev{8,10}},\frac{\vev{2,10}\vev{4,6}}{\vev{2,4}\vev{6,10}}\right\}
\end{eqnarray}
\\

Avoiding to choose both $x$ and $x^{-1}$, our 10 \XX-coordinates are:
\beqa
v_1= \frac{\vev{13}\vev{57}}{\vev{17}\vev{35}} ,\quad v_3= \frac{\vev{35}\vev{79}}{\vev{39}\vev{57}} ,\quad
 v_5= \frac{\vev{19}\vev{57}}{\vev{15}\vev{79}} ,\quad v_7=\frac{\vev{13}\vev{79}}{\vev{37}\vev{19}}  ,\quad
  v_9&=& \frac{\vev{19}\vev{35}}{\vev{13}\vev{59}} \nn\\
v_2=\frac{\vev{24}\vev{68}}{\vev{28}\vev{46}} ,\, v_4= \frac{\vev{46}\vev{8,10}}{\vev{4,10}\vev{68}} ,\, v_6= \frac{\vev{2,10}\vev{6,8}}{\vev{2,6}\vev{8,10}} ,\,  v_8=\frac{\vev{24}\vev{8,10}}{\vev{48}\vev{2,10}}  ,\, v_{10}&=&\frac{\vev{2,10}\vev{4,6}}{\vev{2,4}\vev{6,10}}.\nn\\
\eeqa

Turning back to cross ratios, the condition (\ref{ijcondition}) only allows the  nontrivial cross ratios $u_{i,i+4}$ for $i=1,..10$ with indices taken module 10. Naming $u_i=u_{i,i+4}$ (as in \cite{Heslop:2010kq}), we find again that all nontrivial cross ratios are the same simple function (\ref{simplefunction}) of cluster \XX-coordinates, bijectly  relating them. 

\subsection{$n=$12}\label{n=12}

The $A_3\times A_3$ cluster algebra is of rank 6 and there are 18  cluster variables (\ref{Aalgebra}). Each $A_3$ sub-algebra contains $C_4=14$ clusters (number of triangulations of an hexagon), and it would  be cumbersome to show all of them here in order to calculate their \XX-coordinates. Hence, we appeal to  \cite{Golden:2013xva}, which found 15 \XX-coordinates for an  $A_3$ cluster algebra. Here we notice remarkable mismatch between the number of cross ratios (18) used in the $n=$12 two-loop remainder function in \cite{Heslop:2010kq}  and \XX-coordinates (30). Similarly, in \cite{Golden:2013xva}, the authors found that  at $n=6$, in 4D kinematics, only  9 out of 15 \XX-coordinates of a single $A_3$ cluster algebra were related to cross ratios of the motivic amplitudes. These \XX-coordinates can be obtained from triangulations of the hexagon (figure \ref{hexagon4D}(a)). We report below the 9 selected \XX-coordinates found in \cite{Golden:2013xva}:

\begin{figure}[h!] 
\begin{center} 
\setlength{\unitlength}{2.8pt} 
\begin{picture}(40,40)(50,0) 
\thicklines 
  \multiput(0,15)(30,0){2}{\line(0,1){15}} 
  \multiput(0,30)(15,-22.5){2}{\line(2,1){15}} 
  \multiput(15,7.5)(15,22.5){2}{\line(-2,1){15}} 
 
  \multiput(15,7.5)(0,30){2}{\circle*{1}} 
  \multiput(0,15)(0,15){2}{\circle*{1}} 
  \multiput(30,15)(0,15){2}{\circle*{1}}

\put(15,5){\makebox(0,0){$6$}} 
\put(-4,15){\makebox(0,0){$1$}} 
\put(-4,30){\makebox(0,0){$2$}} 
\put(15,40){\makebox(0,0){$3$}} 
\put(34,30){\makebox(0,0){$4$}}
\put(34,15){\makebox(0,0){$5$}}
\put(15,0){\makebox(0,0){$(a)$}}

\end{picture}

\begin{picture}(0,0)(15,-5) 
\thicklines 
  \multiput(0,15)(30,0){2}{\line(0,1){15}} 
  \multiput(0,30)(15,-22.5){2}{\line(2,1){15}} 
  \multiput(15,7.5)(15,22.5){2}{\line(-2,1){15}} 
 
  \multiput(15,7.5)(0,30){2}{\circle*{1}} 
  \multiput(0,15)(0,15){2}{\circle*{1}} 
  \multiput(30,15)(0,15){2}{\circle*{1}}

\put(15,5){\makebox(0,0){$11$}} 
\put(-4,15){\makebox(0,0){$1$}} 
\put(-4,30){\makebox(0,0){$3$}} 
\put(15,40){\makebox(0,0){$5$}} 
\put(34,30){\makebox(0,0){$7$}}
\put(34,15){\makebox(0,0){$9$}}
\put(15,0){\makebox(0,0){$(b)$}}

\end{picture}
\begin{picture}(0,0)(-35,-5) 
\thicklines 
  \multiput(0,15)(30,0){2}{\line(0,1){15}} 
  \multiput(0,30)(15,-22.5){2}{\line(2,1){15}} 
  \multiput(15,7.5)(15,22.5){2}{\line(-2,1){15}} 
 
  \multiput(15,7.5)(0,30){2}{\circle*{1}} 
  \multiput(0,15)(0,15){2}{\circle*{1}} 
  \multiput(30,15)(0,15){2}{\circle*{1}}

\put(15,5){\makebox(0,0){$12$}} 
\put(-4,15){\makebox(0,0){$2$}} 
\put(-4,30){\makebox(0,0){$4$}} 
\put(15,40){\makebox(0,0){$6$}} 
\put(34,30){\makebox(0,0){$8$}}
\put(34,15){\makebox(0,0){$10$}}
\put(15,0){\makebox(0,0){$(c)$}}

\end{picture}
\end{center} 
\caption{ In 4D kinematics, $n=$6, \XX-coordinates are obtained from triangulation of the hexagon (a).  In 2D kinematics, $n=$12, \XX-coordinates are obtained from triangulations of hexagons (b) and (c).}
\label{hexagon4D}
\end{figure}
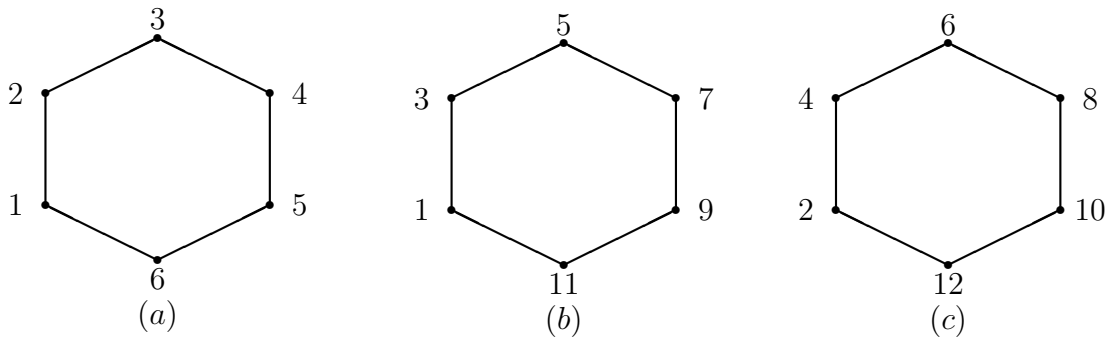  

\begin{align}
\label{eq:nineratios}
& \frac{\ket{23} \ket{56}}{\ket{35}\ket{26}},&
 & \frac{\ket{61} \ket{34}}{\ket{13}\ket{64}},&
 & \frac{\ket{45}\ket{12}}{\ket{15} \ket{24}},
\nonumber\\
 & \frac{\ket{12}\ket{34}}{\ket{14}\ket{23}},&
 & \frac{\ket{56}\ket{12}}{\ket{25}\ket{16}},&
 & \frac{\ket{34}\ket{56}}{\ket{36}\ket{45}},
\\
 & \frac{\ket{45}\ket{16}}{\ket{14}\ket{56}},&
 & \frac{\ket{23}\ket{45}}{\ket{25}\ket{34}},&
& \frac{\ket{61}\ket{23}}{\ket{63}\ket{12}}
\nonumber
\end{align}
We made above a convenient choice between some \XX-coordinates and their inverse which will be later clear. Will the above  selected \XX-coordinates also play a role in special 2D kinematics? At $n=$12 in special 2D kinematics,  we have two copies of $A_3$ cluster algebra and therefore two copies of the above \XX-coordinates, one with all odd particles and the other with all even particles (figures \ref{hexagon4D} (b) and (c)). 
They are:

\begin{align}
\begin{aligned}
v_{5,11}=\frac{\ket{57} \ket{11,1}}{\ket{7,11}\ket{51}}\\ v_{4,10}= \frac{\ket{4,6} \ket{10,12}}{\ket{6,10}\ket{4,12}}
\end{aligned} &\;,&
\begin{aligned} 
 v_{17}= \frac{\ket{13} \ket{79}}{\ket{37}\ket{19}}\\ v_{6,12}=\frac{\ket{12,2} \ket{68}}{\ket{26}\ket{12,8}}
 \end{aligned}&\;,&
 \begin{aligned}
 v_{39}= \frac{\ket{35}\ket{9,11}}{\ket{3,11} \ket{59}}\\ v_{2,8}=\frac{\ket{24}\ket{8,10}}{\ket{4,10} \ket{48}}
\end{aligned}\nn\\
\begin{aligned} v_{37}= \frac{\ket{35}\ket{79}}{\ket{39}\ket{57}}\\ v_{26}=\frac{\ket{24}\ket{68}}{\ket{28}\ket{46}}\end{aligned}&\;,&
  \begin{aligned} v_{11,3}=\frac{\ket{11,1}\ket{35}}{\ket{5,11}\ket{31}}\\ v_{10,2}=\frac{\ket{10,12}\ket{24}}{\ket{4,10}\ket{2,12}}\end{aligned}&\;,&
  \begin{aligned}v_{7,11}=\frac{\ket{79}\ket{11,1}}{\ket{71}\ket{9,11}}\\ v_{6,10}=\frac{\ket{68}\ket{10,12}}{\ket{6,12}\ket{8,10}}\end{aligned}
  \nonumber\\
 \begin{aligned}v_{91}=\frac{\ket{9,11}\ket{13}}{\ket{39}\ket{1,11}}\\ v_{8,12}=\frac{\ket{8,10}\ket{2,12}}{\ket{28}\ket{10,12}}\end{aligned}&\;,&
  \begin{aligned}v_{59}=\frac{\ket{57}\ket{9,11}}{\ket{5,11}\ket{79}}\\ v_{48}=\frac{\ket{46}\ket{8,10}}{\ket{4,10}\ket{68}}\end{aligned}&\;,&
 \begin{aligned}v_{15}=\frac{\ket{13}\ket{57}}{\ket{17}\ket{35}}\\ v_{12,4}=\frac{\ket{12,2}\ket{46}}{\ket{12,6}\ket{24}}\end{aligned}
 \label{n12xcoords}
\end{align}

Once more we find that the same direct relation of the above \XX coordinates to the preferred set of cross-ratios (\ref{simplefunction}) chosen in \cite{Heslop:2010kq} for describing $\mathcal{R}_{12}^{(2)}$ in special 2D kinematics. Rewriting the expression  (\ref{crossratios})  for cross ratios using Pl\"ucker relation:
\beq
\label{crossratioXcoord}
u_{ij}=\frac{1}{1+\frac{\ket{i, i+2}\ket{j,j+2}}{\ket{i,j+2}\ket{i+2, j}}}
\eeq   
which implies that the  18 nontrivial  cross ratios in \cite{Heslop:2010kq} for $n=12$, $u_{i,i+4}$ and $u_{i,i+6}$, $i=1,..12$, (ciclic indices) are simple expressions  (\ref{simplefunction}) of the above \XX-coordinates of $A_3\times A_3$ cluster algebra:
\beq
u_{i,i+4}=\frac{1}{1+v_{i,i+4}}\quad {\rm and}\quad u_{i,i+6}=\frac{1}{1+v_{i,i+6}}
\label{n12crossratios}
\eeq   
for $i=1..12$ and indices module 12.

Intricately, 12 \XX-coordinates made no appearance in relation to cross ratios, which led us to make a more careful look at the 18 \XX-coordinates that appeared in equation (\ref{n12crossratios}). For simplicity, we will work with only one copy  $A_3$ of $A_3\times A_3$ cluster algebra. 

\begin{figure}[ht] 
\includegraphics[width=15cm]{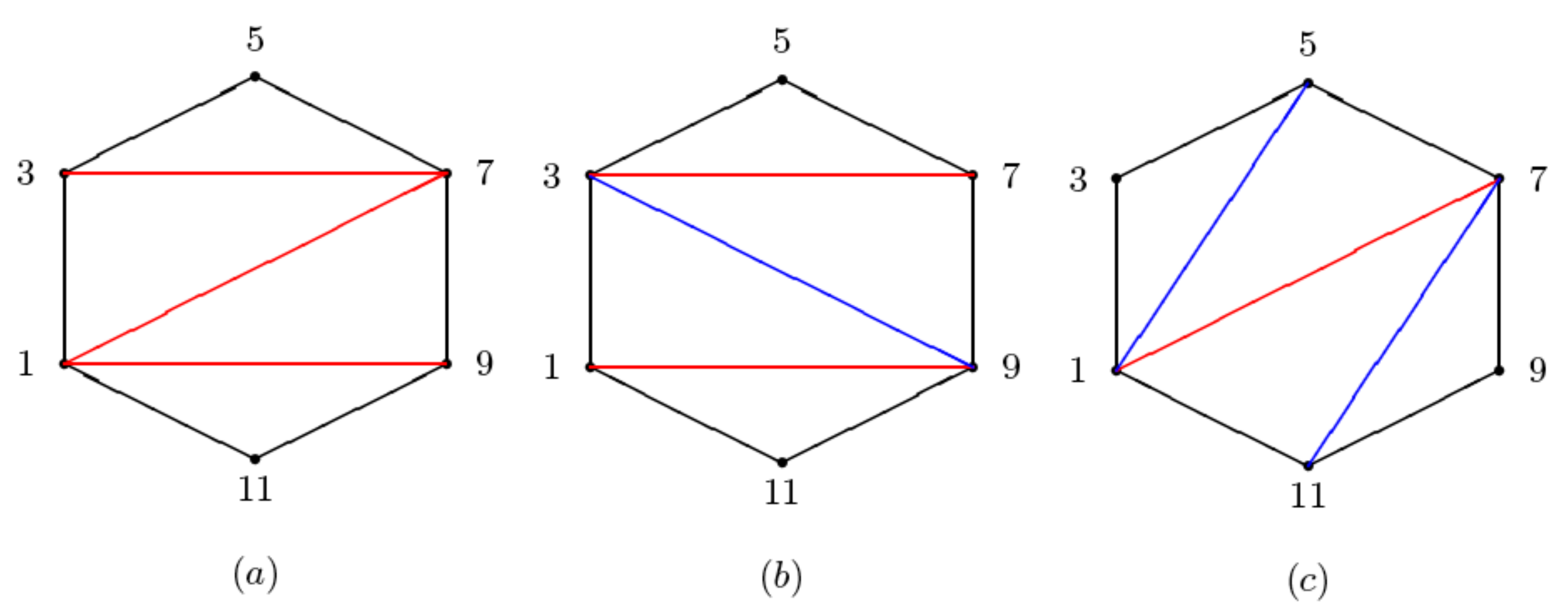}
\caption{3 out of 6 snake  triangulations in hexagon. Triangulation (b) is a mutation exchange of diagonal $\overline { 17 }$ in (a) and triangulation (c)  is two mutations away from (a) by exchanging at steps diagonals $\overline{37}$ and $\overline{19}$ by $\overline{15}$ and $\overline{7\,11}$ respectively. } 
\label{snakes in hexagon} 
\end{figure}

We start by noticing that \XX-coordinates in (\ref{n12xcoords}) are originated from clusters associated to snake triangulations. According to expression (\ref{ijcondition}), looking at figure \ref{snakes in hexagon} we see that the \XX-coordinates corresponding to diagonals in snake triangulations (a), (b) and (c) are
\begin{align}
&{\rm hexagon\, (a)}& & {\rm  hexagon \,(b)} & &{\rm hexagon\, (c )}\quad\\ \nn
&\overline{17}  \rightarrow  v_{17} & &\overline{39}  \rightarrow   v_{1,7}^{-1} & & \overline{17} \rightarrow  v_{5,11}^{-1}\\ \nn
&\overline{19}  \rightarrow  v_{7,11}^{-1}& &     \overline{19}  \rightarrow  v_{19}    & &\overline{7\,11}  \rightarrow  v_{7,11}\\ \nn
&\overline{37}  \rightarrow  v_{15}^{-1}& &  \overline{37}  \rightarrow  v_{37}  &&  \overline{15} \rightarrow  v_{15}
\end{align}

Continuing checking with the three other remaining snake triangulations (not shown here) or simply rotating the indices, one can realize that all 9 \XX-coordinates of odd parity that are directly related to cross ratios are part of clusters associated to snake triangulations. We will call them snake clusters. 

 Notice that these snake triangulations hold a remarkable property of projecting  $A_1$ and $A_1\times A_1$ cluster subalgebras  out of  $A_3$ cluster algebra. These cluster  sub-algebras correspond to lines and quadrilateral faces in the $A_3$ Stasheff polytope \cite{Golden:2013xva} passing through six snake clusters vertices in the equator of the polytope. We see that by noticing in figure \ref{snakes in hexagon} that the snake triangulation in (a) mutates to the snake triangulation in (b) by exchanging diagonal $\overline{17}$ to $\overline{39}$ which exchanges $v_{17}$ by its inverse. But in order to change from  the snake triangulation in (a) to (c)  two mutations are necessary, and these two mutations can be done in any order, characterizing a $A_1\times A_1$ quadrilateral face:

\begin{equation}
\label{eq:square}
\begin{gathered}
\begin{xy} 0;<1pt,0pt>:<0pt,-1pt>::
(180,0) *+{\framebox[30ex]{snake (c ) $\{v_{7,11},v_{1,5},\ldots\}$}}  ="0",
(180,50) *+{\framebox[30ex]{$\{v_{7,11},v_{1,5}^{-1},\ldots\}$}}  ="1",
(0,50) *+{\framebox[30ex]{snake (a) $\{v_{7,11}^{-1},v_{1,5}^{-1},\ldots\}$}} ="2",
(0,0) *+{\framebox[30ex]{$\{v_{7,11}^{-1} ,v_{1,5},\ldots\}$}} ="3",
"0";"1" **\dir{-};
"3";"0" **\dir{-};
"1";"2" **\dir{-};
"2";"3" **\dir{-};
\end{xy}\end{gathered},
\end{equation}
 \\
 
 The transitions between snake triangulations (a) and (c)  can be seen as the freezing of diagonal $\overline{17}$, leading to the formation of two quadrilaterals $\overline{1357}$ and $\overline{7\,9\,11\,1}$ each representing an $A_1$ cluster sub-algebra.  $A_1\times A_1$ cluster algebra 
 has only two \XX-coordinates and in diagram (\ref{eq:square}) they are $v_{7,11}$ and $v_{1,5}$. 
 
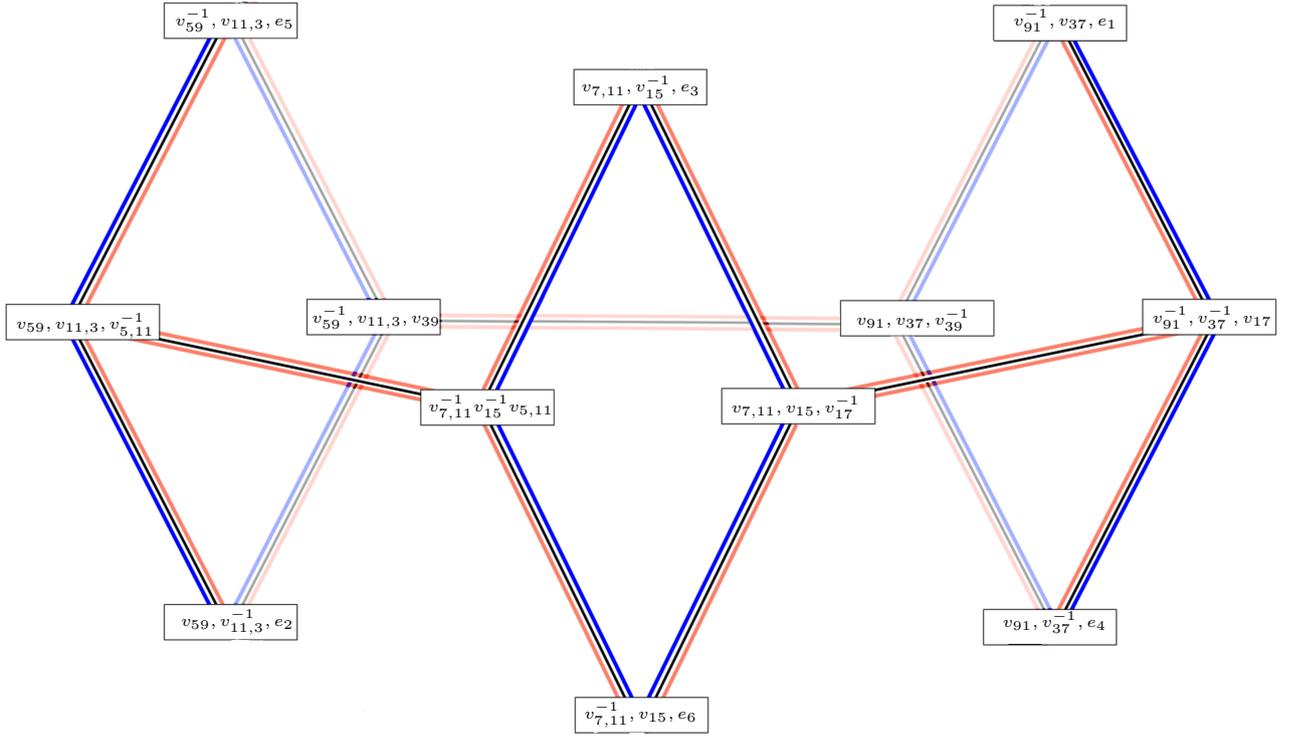
\begin{figure}
  \begin{picture}(200,250)
  \put(-16,0){\includegraphics[width=17.0truecm]{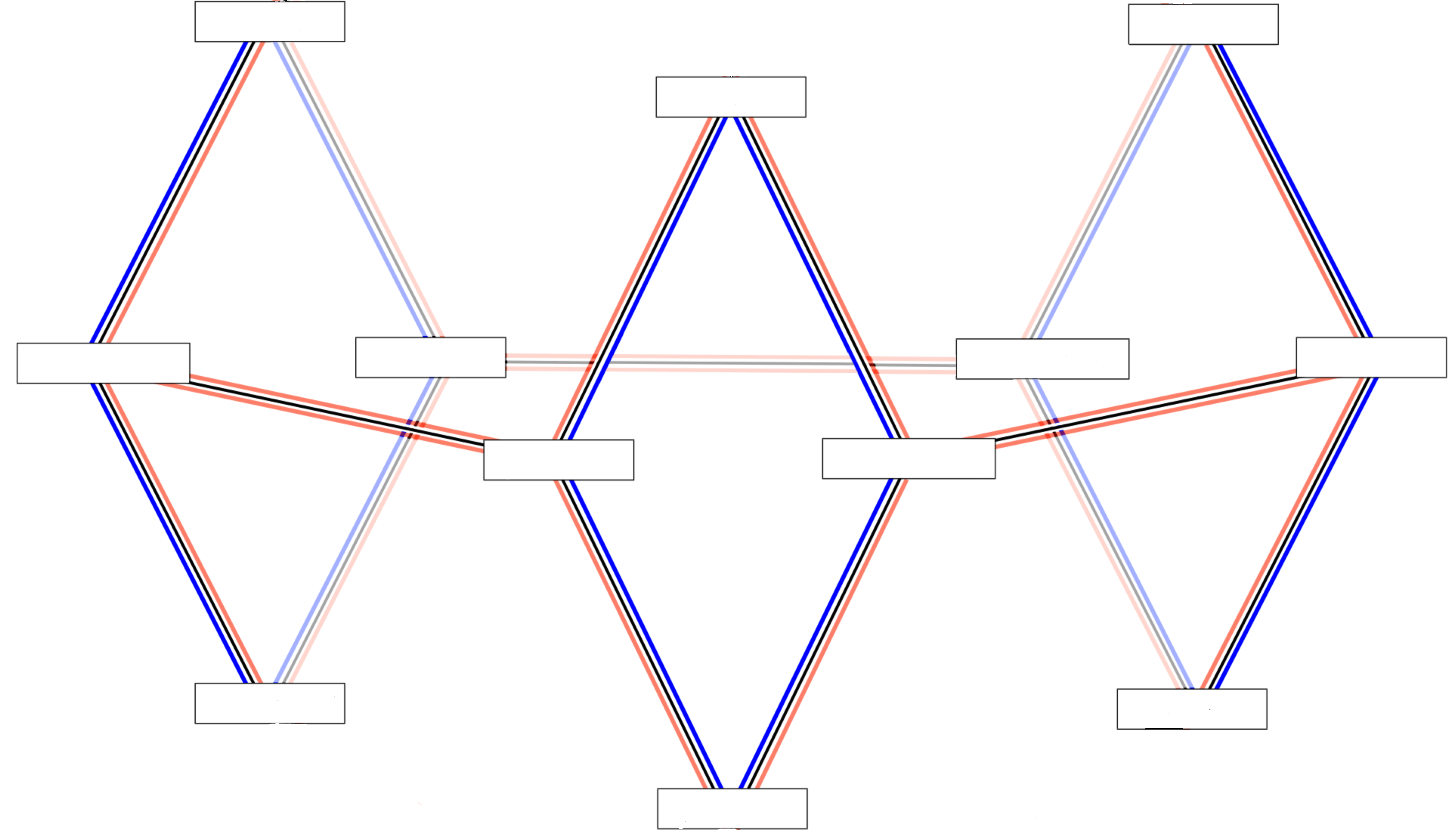}}
\put(51,35){\makebox(50,15){\tiny $v_{59},v_{11,3}^{-1},e_2 $}}
\put(-6,147){\makebox(50,15){\tiny $v_{59},v_{11,3},v_{5,11}^{-1} $}}
\put(103,150){\makebox(50,15){\tiny $v_{59}^{-1},v_{11,3},v_{39} $}}
\put(50,262){\makebox(50,15){\tiny $v_{59}^{-1},v_{11,3},e_5 $}}
\put(304,149){\makebox(50,15){\tiny $v_{91},v_{37},v_{39}^{-1} $}}
\put(146,116){\makebox(50,15){\tiny $v_{7,11}^{-1}v_{15}^{-1}v_{5,11}$}}
\put(260,116){\makebox(50,15){\tiny $v_{7,11},v_{15},v_{17}^{-1}$}}
\put(202,0){\makebox(50,15){\tiny $v_{7,11}^{-1},v_{15},e_6 $}}
\put(202,237){\makebox(50,15){\tiny $v_{7,11},v_{15}^{-1},e_3$}}
\put(416,150){\makebox(50,15){\tiny $v_{91}^{-1},v_{37}^{-1},v_{17} $}}
\put(361,262){\makebox(50,15){\tiny $v_{91}^{-1},v_{37},e_1 $}}
\put(357,35){\makebox(50,15){\tiny $v_{91},v_{37}^{-1},e_4 $}}
  \end{picture}
  \caption{Part of the Stasheff polytope from $A_3$ cluster algebra, obtained mutation exchanges between the 6 snake clusters vertices in the equator of the picture.}
  \label{partial Stasheff}
\end{figure}

 Figure (\ref{partial Stasheff}) shows the partial Stasheff polytope obtained transitioning between snake triangulations (or snake clusters). Each vertex represent a cluster defined by its 3 \XX-coordinates. Edges connecting vertices represent a mutation exchange between clusters. Vertices outside the equator do not correspond to snake triangulations and they are parametrized by some of the six $\{e_i\}$ \XX-coordinates not mentioned in (\ref{n12xcoords}).  
 
\section{General $n$ external particles}
\label{general n}

Expression (\ref{crossratioXcoord})  suggest us that the standard basis of cross ratios  $u_{ij}$  (i and j of same parity) in special 2D kinematics (\ref{crossratios}) are expression of \XX-coordinates $v_{ij}$ of the form $u_{ij}^{-1}=1+v_{ij}$,  once we are able to show that there is a set of \XX-coordinates given by 
\beq
v_{ij}= \frac{\ket{i, i+2}\ket{j,j+2}}{\ket{i,j+2}\ket{i+2, j}}\quad,    4\leq j-i \leq n-4 , \quad (j-i)\; {\rm even}
\label{Xcoords}
\eeq

We saw before that for $n=12$ there is a larger number of \XX-coordinates than standard cross ratios and that some \XX-cooordinates  are like (\ref{Xcoords}) , directly related to cross ratios (\ref{crossratios}), setting them apart from the remaining \XX-coordinates. We show below that the cluster structure offers some explanation, highlighting them in the cluster structure.  

Most of the concepts have been already presented in subsection (\ref{n=12}). Here we will put it in general terms. Again, for simplification, we will look only at one $A_{n/2-3}$ cluster sub-algebra of the entire $A_{n/2-3}\times A_{n/2-3}$ cluster algebra.

\begin{figure}[ht] 
\begin{center} 

%
%

%
%
\includegraphics[width=7cm]{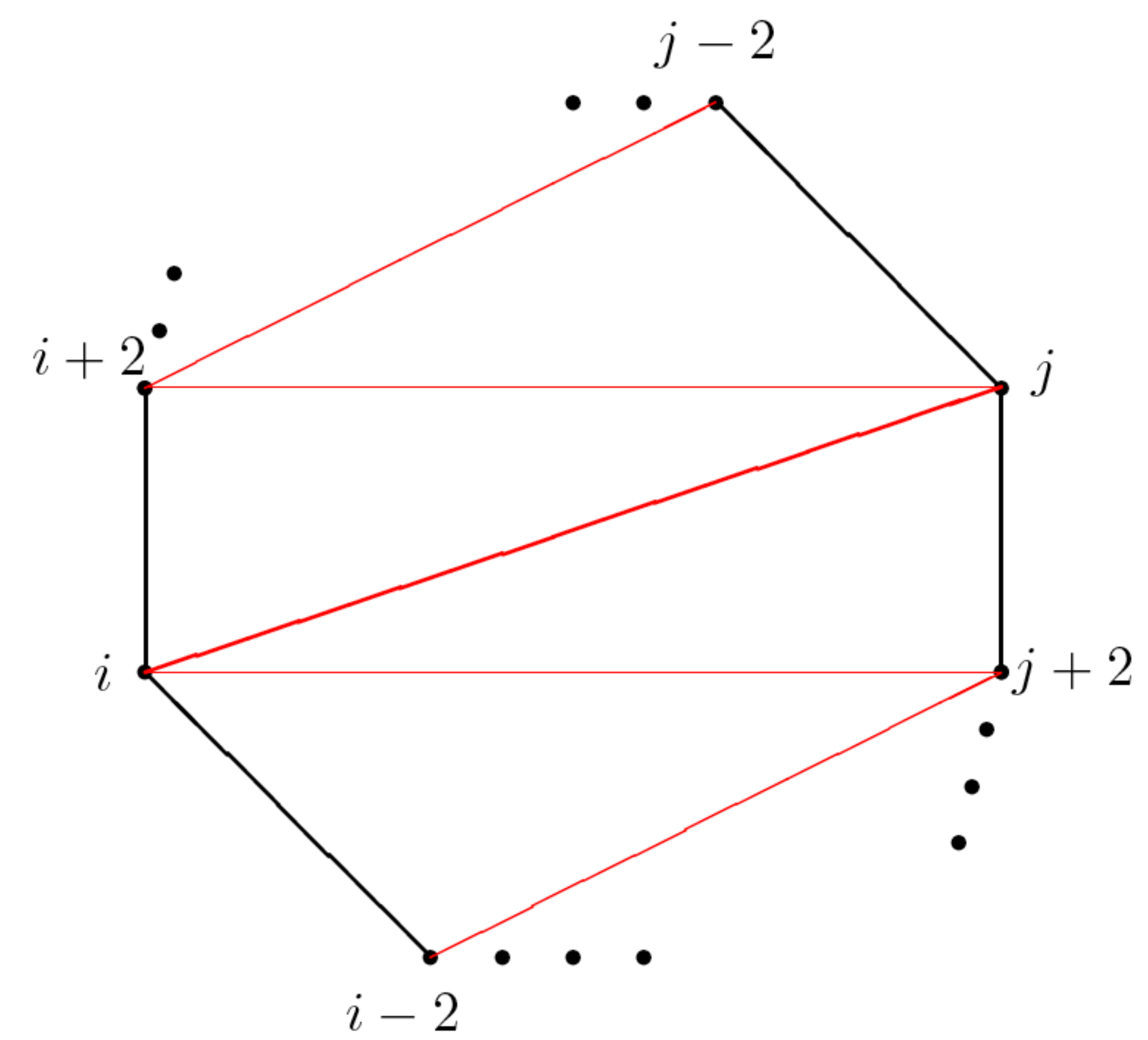}
\end{center} 
\caption{$\overline{ij}$ diagonal in a ``snake''  triangulation of an n-gon} 
\label{fig:n-snake} 
\end{figure}

\begin{figure}[ht] 
\begin{center} 

\includegraphics[width=13cm]{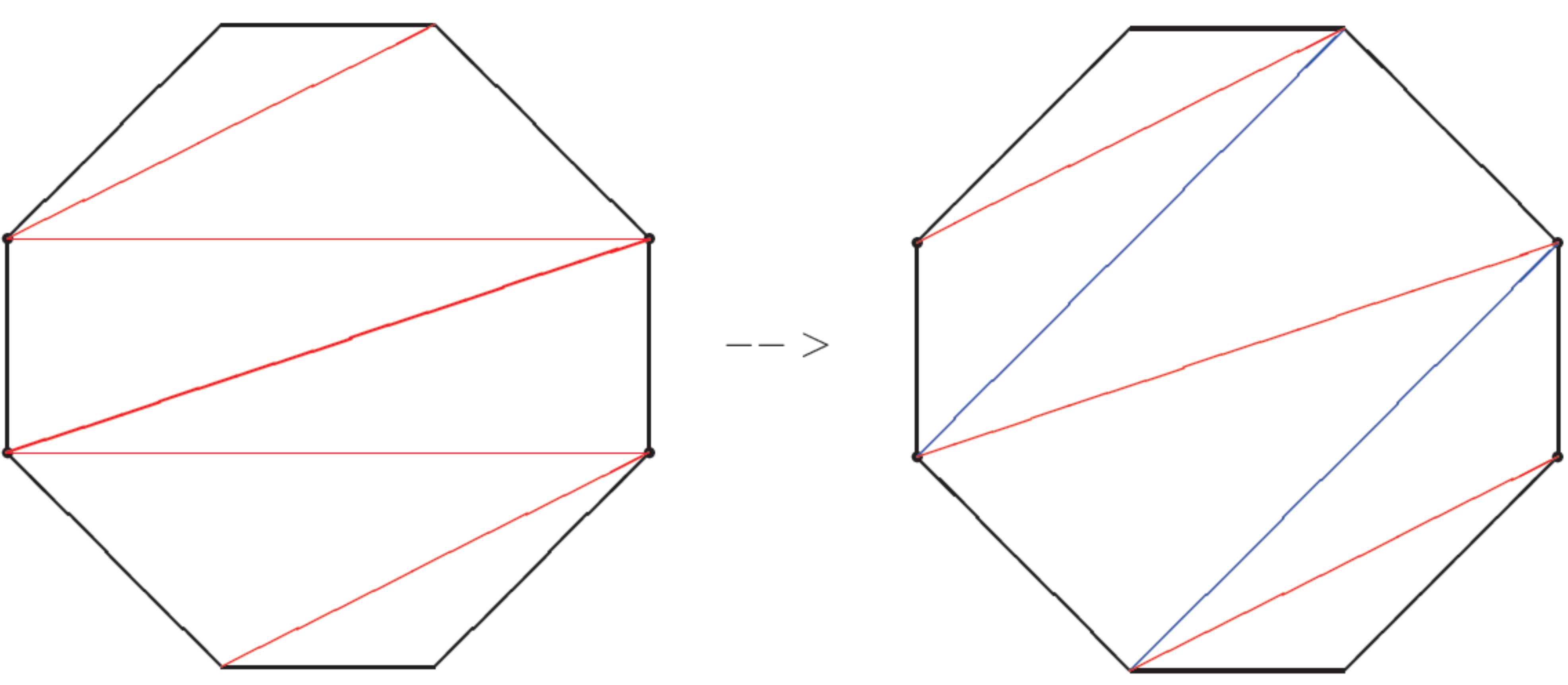}
\end{center} 
\caption{Exchanging  ``snake''  triangulations by two mutations of alternating ``zig'' diagonals.} 
\label{snake exchange} 
\end{figure}

In order to show (\ref{Xcoords}), we start by noticing that any $\overline{ij}$ diagonal can be part of a  zig-zag or snake triangulation \cite{FominZ} of a polygon of $n/2$ sides. From a initial  snake triangulation, all other snake triangulations can be obtained by step by step mutation exchange of diagonals without common vertices in the initial triangulation (Fig.\ref{snake exchange}) (mutates only the zig and skips the zag diagonals). In this way, any diagonal $\overline{ij}$ is part of a unique ``snake'' triangulation, crossing quadrilateral $\{i, i+2, j,j+2\}$ (Fig. \ref{fig:n-snake}).  The \XX-coordinates  of a diagonal $\overline{ij}$ in such  snake triangulation are of the form (\ref{Xcoords}). The total number of such \XX-coordinates is the same as the total number of \A-coordinates and the number of cross-ratios (\ref{crossratios}).

\begin{figure}
\centering
\includegraphics[width=5.0truecm]{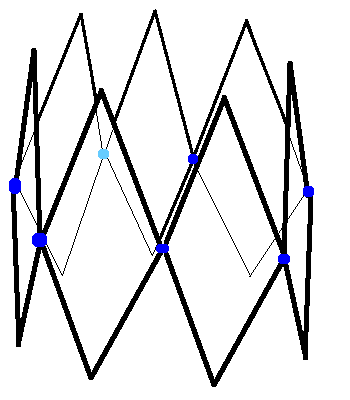}
\caption{$A_4$ partial Stasheff polytope related to snake triangulations of an heptagon. Vertices in blue correspond to snake clusters. Seven quadrilaterals corresponding to $A_1\times A_1$ cluster subalgebra.}
\label{crown}
\end{figure}

The total number of snake triangulations of a $n/2$-gon is $n/2$. In the corresponding Stasheff polytope, these $n/2$ vertices will be separated by hypercubes of dimension $p=\Big[\frac{n/2-3}{2}\Big]$ and $q=n/2-3 -p$, where brackets mean integer part. These hypercubes represent $A_1 \times A_1 \times \dots A_1$ cluster sub-algebras of $A_{n/2-3}$ cluster algebra. These cluster sub-algebras $A_1^p$ and $A_1^q$ can be seen through  mutations from one snake triangulation vertex to the next snake triangulation vertex in the Stasheff polytope. Departing from a snake triangulation (figure \ref{snake exchange}),  the two simplest set of mutations to get to another snake configuration are mutating  all zig or all zag diagonals  while keeping the remaining zag or zig diagonals unchanged. Keeping diagonals unchanged can be seen as the freezing of the \A-coodinates that they represent and it leads to the splitting of the polygon in several quadrilaterals, each one corresponding to a $A_1$ cluster sub-algebra.  In figure \ref{crown}, we exemplify a $n=14$ case corresponding to a partial  $A_4$  Stasheff polytope made of seven snake cluster vertices and seven quadrilaterals  (p=q=2). Such structure of snake clusters and hypercubes in the Stasheff polytope we name necklace of hypercube beads, or hypercube necklace, as figure \ref{crown} suggests. Freezing of different diagonals in the snake triangulation of a $n/2$-gon generate other type A cluster sub-algebras and other geometrical pictures with one snake cluster vertex in the Stasheff polytope .   

Hypercube necklaces in Stasheff polytopes are important in counting  quadrilateral faces composed  of the \XX-coordinates from snake clusters selected by the amplitude. Quadrilateral faces ($A_1\times A_1$ cluster sub-algebra) of amplitude selected  \XX-coordinates were found \cite{Golden:2013lha} to be related to obstructions terms of Bloch groups $B_2\times B_2$  which do not allow the MHV two-loop motivic amplitude to be written in terms of sum of classical 4-logarithms. In 4D kinematics, no obstruction were found at $n = 6$ ($A_3$ cluster algebra), but at $n=7$  ($E_6$ cluster algebra) it was found that the  obstruction is nontrivial  and depends on only 42 quadrilateral faces, out of 1785 existent in $E_6$ Stasheff polytope.
 
\section{Conclusion}
\label{conclusion}
Special 2D kinematics offers an interesting arena to study the cluster structure of the configuration space of a n-particle amplitude. Following the steps of \cite{Golden:2013xva}, we notice the existence of a preferential choice of kinematic variables (cross-ratios) that preserve the cluster structure of the configuration space of external particles and simultaneously leading to simple amplitude expressions  (\cite{Heslop:2010kq}) in special 2D kinematics. The cluster algebras in 2D kinematics are of $A_i$ type. $A_i$ cluster algebras also appear in 4D kinematics either as an entire cluster algebra or as cluster sub-algebras and their importance remains in relation to the choice of nontrivial cross ratios. 

In the present work, we showed the existence of special \XX-coordinates among all \XX-coordinates of a cluster algebra of type $A_n$. They are \XX-coordinates  of snake clusters,  those whose associated polygon triangulation  are snakes. In an $A_n$ cluster algebra, there are $n+3$ snake clusters. In Stasheff polytope these snake clusters are opposite vertices of hypercubes of dimension p and q (a line if p=1 or quadrilateral face if p=2), such that $p=[n/2]$ and $p+q=n$. These snake cluster vertices connected by hypercubes form a necklace in the $A_n$ Stasheff polytope with hypercubes as beads. We call it a hypercube necklace. Such structure contains many $A_1 \times A_1$ quadrilateral faces made of pairs of snake cluster \XX-coordinates  that can be important in relation to obstruction terms found in \cite{Golden:2013xva}. For $E_6$ cluster algebra, by freezing cluster variables in the $E_6$ quiver we checked the presence of four $A_4$ necklaces and  one $A_5$ necklace summing all together 56 quadrilateral faces, which some may play a special role as arguments in $B_2\times B_2$ obstruction terms found at  $n = 7$  (4D kinematics) amplitude.   
 
 Furthermore, we showed that in special 2D kinematics at any number of external particles, the canonical choice of cross-ratios (\ref{crossratios}) selected by the simple two-loop MHV amplitude expression of \cite{Heslop:2010kq} is directly related to these special \XX-coordinates. 
 
In \cite{Golden:2013xva}, the authors noticed that in  two-loop MHV $n= 6$ amplitude, only 9 out of 15 \XX-coordinates appear in the remainder function $\mathcal{R}_6^{(2)}$. This case is a $A_3$ cluster algebra of 
$\Conf_6(\PP^3)$. In special 2D kinematics $A_3$ cluster algebra appears in $n=12$ amplitude, and we showed that the same  9 \XX-coordinates  out of 15 are selected to appear in the respective two-loop remainder function. The structure of the cluster algebra discussed here have a fundamental role in this mysterious selection. The fact that the same selection took place at 2D and 4D kinematics with different amplitude expressions, reveals the importance of the cluster structure of the configuration space and its preservation at two loops. 

The structures that we unlocked at special 2D kinematics are always $A_n$ type cluster algebra but once $A_n$ type cluster sub-algebras are part of every configuration space and other kinematical limits can lead to appearance of $A_n$ cluster algebra, we believe that this work program may have significant  implications in 4D kinematics scattering amplitudes.  

\section*{Acknowledgements}
The author is indebted to Ruth Britto for stimulating discussions and criticism on the project  and Bo Feng for early discussions on cluster algebra and his encouraging comments in the early stages of the project. The author also would like to thank Gregory Korchemsky, Simon Caron-Huot and Song He for helpful comments.
 The author is supported by CAPES (Brazilian research agency) postdoctoral fellowship.

\end{document}